\documentclass[manuscript,screen, nonacm]{acmart}
\usepackage{xcolor} 
\usepackage{colortbl} 
\usepackage{multirow} 

\newlength{\blocksize}
\newcommand{\tealblock}[1]{\textcolor{teal}{\rule{#1\blocksize}{0.1cm}}}
\newcommand{\violetblock}[1]{\textcolor{violet}{\rule{#1\blocksize}{0.1cm}}}

\newcolumntype{P}[1]{>{\centering\arraybackslash}p{#1}}

\AtBeginDocument{%
  \providecommand\BibTeX{{%
    \normalfont B\kern-0.5em{\scshape i\kern-0.25em b}\kern-0.8em\TeX}}}

\copyrightyear{2024}
\acmYear{2024}
\setcopyright{rightsretained}
\acmConference[CHI '24]{Proceedings of the CHI Conference on Human Factors in Computing Systems}{May 11--16, 2024}{Honolulu, HI, USA} \acmBooktitle{Proceedings of the CHI Conference on Human Factors in Computing Systems (CHI '24), May 11--16, 2024, Honolulu, HI, USA} \acmDOI{10.1145/3613904.3642118}
\acmISBN{979-8-4007-0330-0/24/05}

\begin{document}

\title[Interconnected eHMIs]{Exploring the Impact of Interconnected External Interfaces in Autonomous Vehicles on Pedestrian Safety and Experience}


\author{Tram Thi Minh Tran}
\email{tram.tran@sydney.edu.au}
\orcid{0000-0002-4958-2465}
\affiliation{Design Lab, Sydney School of Architecture, Design and Planning,
  \institution{The University of Sydney}
  \city{Sydney}
  \state{NSW}
  \country{Australia}
}

\author{Callum Parker}
\email{callum.parker@sydney.edu.au}
\orcid{0000-0002-2173-9213}
\affiliation{Design Lab, Sydney School of Architecture, Design and Planning,
  \institution{The University of Sydney}
  \city{Sydney}
  \state{NSW}
  \country{Australia}
}

\author{Marius Hoggenmüller}
\email{marius.hoggenmuller@sydney.edu.au}
\orcid{0000-0002-8893-5729}
\affiliation{Design Lab, Sydney School of Architecture, Design and Planning,
  \institution{The University of Sydney}
  \city{Sydney}
  \state{NSW}
  \country{Australia}
}

\author{Yiyuan Wang}
\email{yiyuan.wang@sydney.edu.au}
\orcid{0000-0003-2610-1283}
\affiliation{Design Lab, Sydney School of Architecture, Design and Planning,
  \institution{The University of Sydney}
  \city{Sydney}
  \state{NSW}
  \country{Australia}
}

\author{Martin Tomitsch}
\email{Martin.Tomitsch@uts.edu.au}
\orcid{0000-0003-1998-2975}
\affiliation{Transdisciplinary School,
  \institution{University of Technology Sydney}
  \city{Sydney}
  \state{NSW}
  \country{Australia}
}

\renewcommand{\shortauthors}{Tran et al.}

\begin{abstract}
Policymakers advocate for the use of external Human-Machine Interfaces (eHMIs) to allow autonomous vehicles (AVs) to communicate their intentions or status. Nonetheless, scalability concerns in complex traffic scenarios arise, such as potentially increasing pedestrian cognitive load or conveying contradictory signals. Building upon precursory works, our study explores `interconnected eHMIs,' where multiple AV interfaces are interconnected to provide pedestrians with clear and unified information. In a virtual reality study (N=32), we assessed the effectiveness of this concept in improving pedestrian safety and their crossing experience. We compared these results against two conditions: no eHMIs and unconnected eHMIs. Results indicated interconnected eHMIs enhanced safety feelings and encouraged cautious crossings. However, certain design elements, such as the use of the colour red, led to confusion and discomfort. Prior knowledge slightly influenced perceptions of interconnected eHMIs, underscoring the need for refined user education. We conclude with practical implications and future eHMI design research directions.

\end{abstract}



\maketitle
\section{Introduction}

To ensure the seamless integration of autonomous vehicles (AVs) into traffic, it is crucial to address both the driver-vehicle interaction and communication with external road users, especially pedestrians. Historically, non-verbal cues like eye contact and hand gestures facilitated understanding between human drivers and pedestrians~\cite{sucha2017pedestriandriver, gueguen2017stare, ren2016eyecontact}. However, with AVs, this dynamic changes. 
External Human Machine Interfaces (eHMIs) have been developed to bridge this communication gap using modalities like LED light bands~\cite{florentine2016notifications, ford2017communication, dey2020color}, projections~\cite{nguyen2019designing, mercedes2015}, auditory signals~\cite{mahadevan2018communicating, deb2020communicatingfeatures}, and haptic cues~\cite{mahadevan2018communicating}. By displaying the AV's intention or operational states, eHMIs enhance pedestrians' perceived safety~\cite{bockle2017sav2p, deb2018investigating, habibovic2018communicating, declercq2019external, faas2020external}, improve receptivity~\cite{deb2018investigating}, promote trust~\cite{hollander2019overtrust, faas2020longitudinal, kaleefathullah2022external}, and facilitate efficient crossing decisions~\cite{hollander2019investigating}.

However, with the prospect of more AVs being integrated into our urban landscapes in the near future, concerns have arisen about the scalability of eHMIs. In high-density public spaces, pedestrians could face sensory overload from the simultaneous signals of multiple eHMIs, leading to confusion or challenges in interpreting the intended message~\cite{mahadevan2018communicating, robert2019future}. This issue becomes more pronounced in mixed-traffic situations where AVs and human-driven cars coexist. If eHMIs convey crossing information based solely on the actions of one AV, without accounting for the presence of other AVs or manually operated vehicles, they might unintentionally pose risks to pedestrians~\cite{locken2019should}. For instance, consider a scenario where one AV stops and its eHMI indicates it is safe for pedestrians to cross, but the vehicle in the adjacent lane does not stop.

Recognising these challenges, recent research has suggested the idea of interconnected eHMIs. These are systems where eHMIs are networked, providing pedestrians with messages derived from multiple AVs instead of a single one. For example, a vehicle aware of the overall traffic situation might play an audio message saying \textit{`I’m stopping, you can cross'}. ~\cite{colley2020towards}. Other researchers have pointed to the potential of eHMIs that indicate a yielding state only after confirming with nearby vehicles about their stopping intent~\cite{hollander2022take} or those that consider all lanes in their recommendations ~\cite{locken2019should}. Such a unified approach could help reduce information clutter, offering pedestrians more coherent cues. However, the idea of interconnected eHMIs is met with some scepticism. Findings from a study by \citet{colley2020towards} suggest that pedestrians might be hesitant to trust these systems due to their novelty and the ingrained perception of each vehicle as independent.

While there is evident interest in interconnected eHMIs, existing discussions remain largely theoretical, with only an auditory interconnected concept explored in depth by \citet{colley2020towards}. Our research seeks to extend this discourse by conceptualising and evaluating a visual-based interconnected eHMI system. Through a mixed-design VR simulation study (N=32), we delve into three eHMI conditions—interconnected, unconnected, and none—with the objective to determine their effects on pedestrian safety, cognitive load, and trust. Importantly, we further investigated how participants' prior knowledge of interconnected eHMIs might influence these outcomes. 

Our study makes two primary contributions: (1) We conceptualise and empirically evaluate visual-based interconnected eHMIs, focusing on pedestrian safety, cognitive load, and trust. (2) We shed light on the pedestrian's learning curve and the educational prerequisites for the proposed system.

\section{Related Work}

\subsection{Pedestrian-Vehicle Communication in the Era of Autonomous Vehicles}

Pedestrians have historically depended on implicit cues from vehicles, such as distance, speed, and deceleration, to determine when it is safe to cross the street~\cite{rasouli2017agreeing, dey2017role, lee2021road}. This is still the case with AVs, as demonstrated by several field studies employing the ghost driver method~\cite{rothenbucher2016ghost, moore2019case}. However, in situations where vehicle movement is ambiguous, like in slow-moving traffic, pedestrians tend to seek explicit cues from drivers. As we transition to a world with a greater presence of AVs, the need for these cues does not disappear; rather, it evolves. The emergence of eHMIs now facilitates effective communication between AVs and pedestrians in their shared environments.

In recent years, there has been a marked increase in the variety and sophistication of eHMI concepts~\cite{rouchitsas2019external, dey2020taming}. Initially, eHMIs echoed traditional vehicular signals: brake lights indicating deceleration~\cite{petzoldt2018brake, monzel2021brake}, and marker lamps indicating the activation of an automated driving system~\cite{sae2019markerlamps}. Yet, technology and design advancements brought forth innovative and intuitive eHMI designs. Some of these simulated human behaviours, using lights to mimic eye contact with pedestrians~\cite{gui2022going, chang2017eyes}, or animations showing a vehicle's directional intent—akin to a human indicating a direction before walking. Some eHMIs even employed advanced technologies like augmented reality to overlay information on real-world scenes~\cite{hesenius2018don, prattico2021comparing, tabone2021towards, tran2022designing} or used dynamic road projections~\cite{umbrellium2017} to provide crucial signals or warnings directly.

However, an observation made during the initial stages of eHMI development is that many concepts were evaluated in isolation, often emphasising singular vehicle-pedestrian interactions~\cite{colley2020unveiling, tran2021review}. While this approach provides a detailed and nuanced understanding of AV-pedestrian interactions, it fails to encompass the broader context of real-world scenarios. A recent scoping review conducted by~\citet{tran2023scoping} highlighted seven key scalability issues of eHMIs. In multi-pedestrian environments, the most pressing issue is the \textit{Clarity of Recipients}, underscoring the necessity for eHMIs to communicate clearly to individual pedestrians in a vehicle's vicinity~\cite{wilbrink2021scaling, dey2021towards, hubner2022external}. In multi-vehicle environments, which are a focus of this paper, the critical challenges that may arise include:

\textit{Information Overload:} In urban environments, pedestrians encounter various information sources, from vehicle cues to environmental prompts like billboards and smartphone distractions~\cite{hollander2020save, lanzer2023interaction}. With the expected rise in eHMIs, pedestrians may face more diverse interfaces and messages from different AVs~\cite{robert2019future, dey2020taming}. The impact of this increased information and visual clutter on pedestrian safety and traffic efficiency is a growing concern in AV-pedestrian research. Some investigated approaches include determining if AVs need to explicitly communicate at all times~\cite{dey2022nonyielding} or aggregating multiple AV communications using infrastructure (e.g., curbstones~\cite{hollander2022take}) or pedestrian devices (e.g., AR glasses~\cite{tran2022designing} and smartphones~\cite{hollander2020save}).

\textit{Safety:} The potential liability of instructing pedestrians to cross, as well as the risk of overtrust, where pedestrians might over-rely on eHMIs and potentially neglect other safety cues, has been highlighted~\cite{andersson2017hello, lagstrom2016avip, dey2020taming, hollander2019investigating}. A study by \citet{mahadevan2019av} further underscores this point, revealing that some participants, after seeing positive eHMI signals, failed to notice other critical risks like distracted drivers in adjacent lanes. This potential for misjudgment is especially alarming in multi-lane crossings due to the direct negative influence of eHMIs on pedestrian safety.

\subsection{V2X Communication and eHMIs}

To address the aforementioned challenges, the domain of eHMIs has begun to overlap with the emerging V2X (Vehicle-to-Everything) communication technologies. In this broader communication network, vehicles can interact with virtually every entity in their vicinity, including pedestrians. Recent studies have presented intriguing V2X integrated solutions. The `Smart Curbstones' concept involves urban infrastructural sensors that communicate safety information to pedestrians through curb-embedded LEDs~\cite{hollander2022take}. Another proposition involves decentralised eHMIs, shifting their functionalities to personal devices~\cite{hollander2020save}. Notably, \citet{tran2022designing} proposed a system wherein pedestrians, equipped with AR glasses, can communicate and negotiate safe crossing opportunities with oncoming AVs.

The interconnected eHMI system is another V2X-based solution. By networking eHMIs, they can derive and communicate messages based on data from multiple AVs rather than just one. This interconnectedness promises a more coherent and consistent message delivery to pedestrians, potentially eliminating the challenges of information overload and message conflicts seen with traditional eHMI designs. While these systems hold great potential, so far, the only study on them has been by \citet{colley2020towards}. Their research focused on auditory interconnected eHMIs, designed to assist visually impaired individuals in navigating traffic. To delve deeper into interconnected eHMIs, our study focus on visual-based design concepts, because the majority of the pedestrian population relies heavily on visual cues when navigating traffic~\cite{dey2020taming}. Recognising this predominance of visual reliance is essential for creating a comprehensive eHMI solution.

A key challenge facing interconnected eHMI research is the prevalent lack of awareness about the connected vehicle technology. A study by \citet{colley2020towards} suggested that gaps in participant knowledge concerning vehicle connectivity could adversely affect their trust. This finding underscores the importance of prior knowledge in shaping pedestrian perceptions of interconnected eHMIs. To address this issue, our study firstly aims to design interconnected eHMIs capable of visualising their connectivity. Secondly, we investigate how prior knowledge of the concept, acquired through pre-instruction, influences user perceptions. We hypothesise that users with foundational understanding of interconnected eHMIs will demonstrate greater trust and more efficient interaction with these systems. This hypothesis aligns with previous research on internal HMIs, which suggests that user instruction can significantly enhance drivers' understanding, interaction performance, acceptance, and trust~\cite{foster2019user, edelmann2020effects}. Furthermore, this investigation contributes to the ongoing discourse of eHMIs regarding the necessity for public education~\cite{dey2020color, faas2021calibrating}. While longitudinal studies related to the learnability and familiarity of eHMIs exist~\cite{faas2020longitudinal, hochman2020pedestrian}, the role of user instruction as a means to bridge initial knowledge gaps remains largely unexplored.

\subsection{Research Questions}
Given the above challenges and gaps identified in the literature, our study addresses two main research questions (RQ):

\textit{RQ1: To what extent will interconnected eHMIs affect pedestrian cognitive load, safety, and trust during multi-lane crossings?} - This question is motivated by the need to compare the performance of interconnected eHMIs with unconnected eHMIs and baseline scenarios where no eHMI is present.
    
\textit{RQ2: How does prior knowledge about the concept of interconnected eHMIs alter pedestrian cognitive load, safety, and trust during multi-lane crossings?} - Given that interconnected eHMIs are a relatively new concept, we are driven to determine if the impact of interconnected eHMIs is contingent on the user's understanding of the technology.

\section{Design Concepts}

\begin{figure*}[ht]
  \centering
  \includegraphics[width=0.65\linewidth]{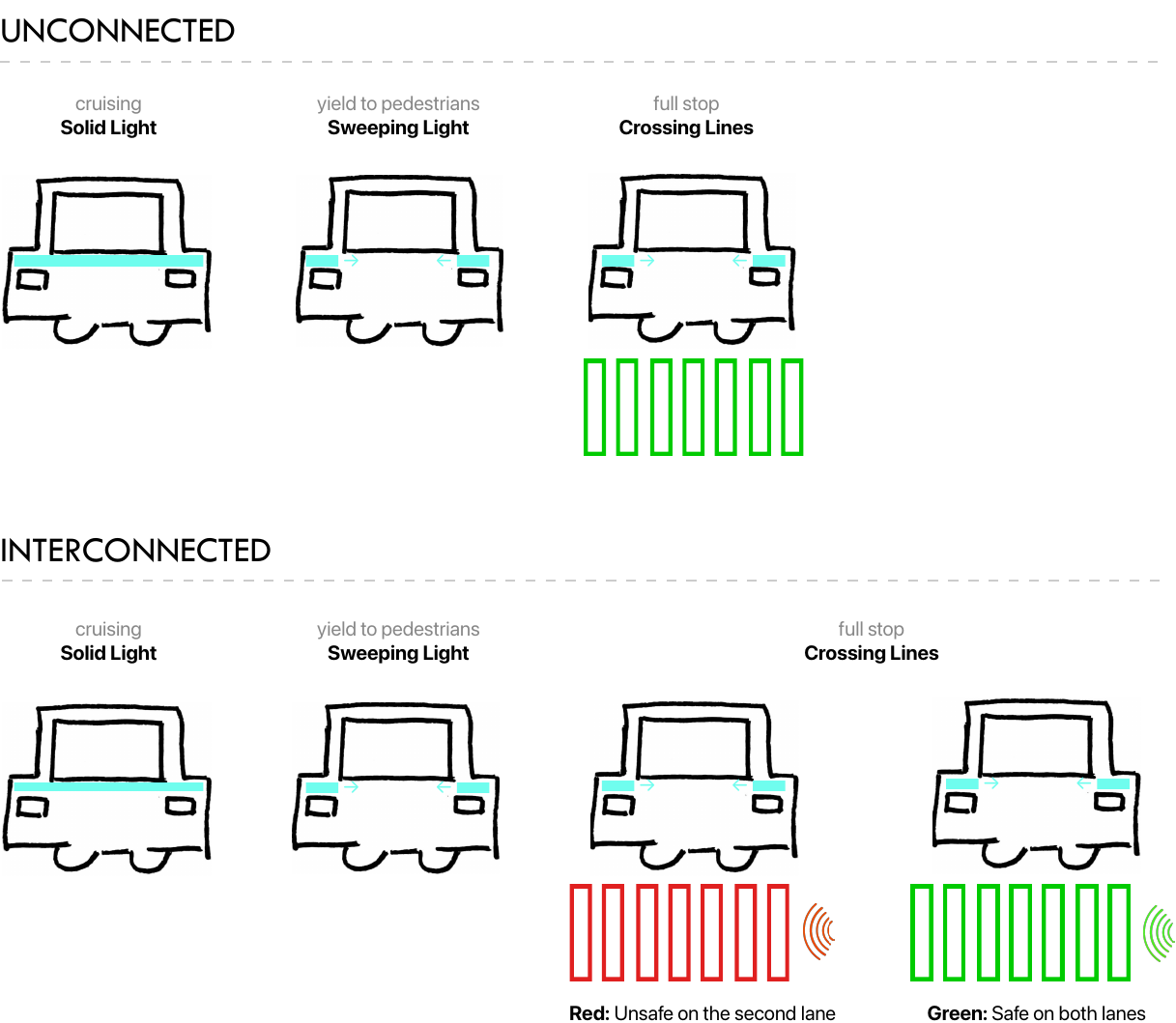}
  \caption{Sketches illustrating eHMI visuals in three distinct AV states: Cruising, Yield to pedestrians, and Full stop. In the unconnected eHMI design (top), a green crossing is projected upon full stop, irrespective of the second lane's conditions. Conversely, in the interconnected eHMI design (bottom), a red crosswalk is displayed when the second lane poses risks, while a green crosswalk appears when both lanes are deemed safe.}
  \Description{The figure depicts sketches that illustrate eHMI visuals for three distinct AV states: Cruising, Yielding to pedestrians, and Full stop. In the unconnected eHMI design, a green crosswalk is projected upon a full stop, regardless of the conditions in the second lane. In contrast, the interconnected eHMI design displays a red crosswalk when the second lane poses risks, and a green crosswalk when both lanes are considered safe. The sketches are categorised into two main sections: the unconnected eHMI design at the top and the interconnected eHMI design at the bottom.}
  \label{fig:conceptcomparison}
\end{figure*}

\subsection{Unconnected eHMI Design} 

Informed by existing literature, three distinct visual representations were selected for the unconnected eHMI, each corresponding to a specific state of the AV (refer to~\autoref{fig:conceptcomparison}). When the vehicle is cruising autonomously, a \textit{solid light band} on its bumper lights up~\cite{habibovic2018communicating, sorokin2019change, dey2020color, loew2022goahead}. As the vehicle prepares to yield to pedestrians (i.e., when applying the brakes), this light band transitions to an \textit{inward-sweeping pattern}. This choice of visualisation was informed by the need for clear differentiation between the actions of giving way and continuing on its path~\cite{dey2022nonyielding}. Upon coming to a complete stop, the vehicle projects \textit{a zebra crossing}, signalling pedestrians about crossing opportunities. This crossing pattern has garnered significant attention in academic research~\cite{tran2022designing, hollander2022take, dey2021towards} and is also a feature of the Mercedes Benz F 015 concept~\cite{mercedes2015}.

The recommended colour for eHMI is typically a blue-green hue, such as cyan or turquoise, chosen for its perceived neutrality, which is not associated with a specific meaning in traffic communication~\cite{pavlo729colors, dey2020color, faas2019color}. As such, the light band is coloured cyan. While \citet{dey2021towards} and \citet{hollander2022take} employed cyan for their zebra crossings, the symbolic representation of a zebra crossing already provides a clear invitation for pedestrians. In this particular context, the distinction between cyan and green is less pronounced. Therefore, we opted for a green crossing pattern, aligning more closely with universal traffic interpretations.

\subsection{Interconnected eHMI Design}

The interconnected eHMI retains many visual cues from its unconnected counterpart, as seen in~\autoref{fig:conceptcomparison}. However, there are several distinct features.

\textit{The Red Crosswalk}:
When the AV comes to a full stop, it projects a red zebra crossing. The red colour is widely recognised for indicating caution and alertness~\cite{wogalter2015color, karyn2014red}. In Australia, red-painted areas on the roads are used to trigger alertness and capture the attention of drivers when they are approaching, for example, a school zone, speed bumps~\cite{drivingtestaustralia2017}, or township entry~\cite{qldtransport2022}. As a result, a combination of red colour and the familiar crosswalk pattern is expected to signal to pedestrians that they can proceed but with increased vigilance. When safety is assured, for instance, when vehicles in adjacent lanes also stop, the crossing transitions from red to green. Thus, the red crosswalk serves a dual function: it helps to mitigate the potential over-reliance of pedestrians on eHMIs for safety decisions and communicates both the intentions of the AV and the broader traffic situation.

\textit{Single AV Communication}: In designs where eHMIs communicate independently, each vehicle projects its own crossing pattern when stopping, even if they stop sequentially in the same lane. A study by \citet{hollander2022take} exemplified such an approach. Our proposed design designates one AV to take charge of the communication and projects the zebra crossing, adopting the `omniscient narrator' concept of the auditory interconnected eHMIs~\cite{colley2020towards}. This centralised approach is intended to minimise visual clutter that could arise from multiple crosswalk projections and aims to provide pedestrians with a single, clear source of information on which to focus their attention. However, understanding that pedestrians might expect to see individual AV signals~\cite{tran2022designing, colley2020towards}, we retain the light band on each AV. Moreover, we synchronise the inward sweeping animation across all AVs, reinforcing the perception of a connected vehicle fleet.

\textit{The Wi-Fi Signal}: Given the unfamiliarity and skepticism towards connected vehicles noted by \citet{colley2020towards}, our design integrates an animated Wi-Fi symbol. This universally recognised sign of connectivity aims to bridge knowledge gaps, fostering immediate understanding of interconnected eHMI operations.

\section{Evaluation Study}

\subsection{Study Design}

We employed a mixed-methods approach using a 2x3 mixed design to explore the influence of knowledge about interconnected eHMIs and differences among three interface conditions, whilst controlling for individual differences.

For the between-subjects factor, participants were randomly allocated to two distinct groups. The \textit{knowledge} group was provided with information about interconnected eHMIs, whereas the \textit{no knowledge} group was not given this information. As for the within-subjects factor, all participants underwent three interface conditions: a baseline without eHMI, an unconnected eHMI, and an interconnected eHMI. Each condition was presented in blocks in a counterbalanced order to avoid learning effects. 

Within each block, participants experienced three different scenarios. The order of these scenarios was randomised among the conditions, ensuring that the sequence of scenarios did not influence participant responses. In total, each participant completed nine crossing trials. In each scenario, an AV on the nearest lane was programmed to yield to the pedestrian 13 seconds after the start of the VR scenario. However, the behaviour of the vehicles on the furthest lane varied to mitigate potential habituation and ensure that participants could experience the design concepts in various traffic situations:

\begin{enumerate}
\item \textit{An AV will yield:} This scenario allows participants to notice the single communication approach of the interconnected eHMIs, represented by only one projected green crosswalk. This is in contrast to the individual communication of both AVs in the unconnected eHMIs, where each AV projects its own green crosswalk, resulting in two separate projections.

\item \textit{A manually-driven vehicle will yield:} This scenario allows participants to observe and understand how the interconnected eHMI concept can respond to non-autonomous vehicles.

\item \textit{No vehicle yielding:} In this scenario, the interconnected eHMI maintains a continuous red signal, despite the presence of sufficiently large time gaps for crossing in the second lane. This cautious approach is likely to occur when the interconnected eHMI system experiences weak connectivity, leading it to default to a more conservative safety protocol. Consequently, this scenario allows us to explore how pedestrians balance interconnected eHMI signals with their own judgement when the communication appears overly cautious.
\end{enumerate}

\subsection{Virtual Reality Simulation}

A VR simulation was selected due to its inherent safety advantages, removing potential risks of direct interactions between pedestrians and AVs in a physical setting~\cite{tran2021review}. In situations with multiple vehicles on multi-lane roads, this safety aspect is noteworthy, as the possibility for collisions might increase (e.g., one collision recorded in a related study~\cite{hollander2022take}). Moreover, VR simulations provide a cost-effective alternative to real-world experiments. The platform's tools enable efficient data collection, offering insights into pedestrian crossing behaviours. 

\begin{figure*}[htbp]
  \centering
  \includegraphics[width=0.6\linewidth]{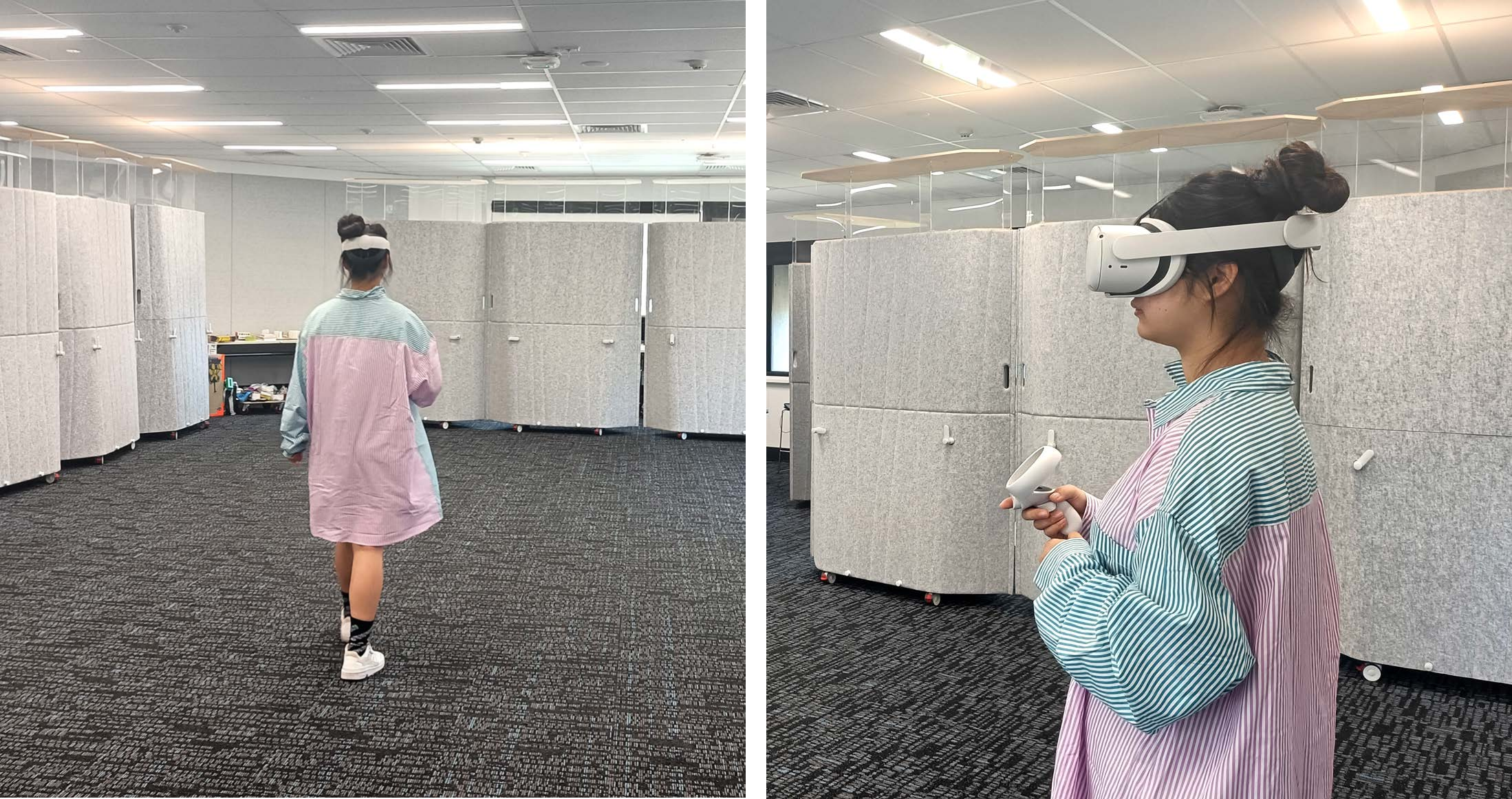}
  \caption{A participant walking with a VR headset on (left). The same participant using a Touch controller to navigate to the next scenario (right).}
  \Description{The figure contains two side-by-side images of a participant in a VR setting. The left image depicts the participant walking in a large space whilst wearing a VR headset. The right image shows the participant using a Touch controller.}
  \label{fig:studysetup}
\end{figure*}

\begin{figure*}[htbp]
  \centering
  \includegraphics[width=0.98\linewidth]{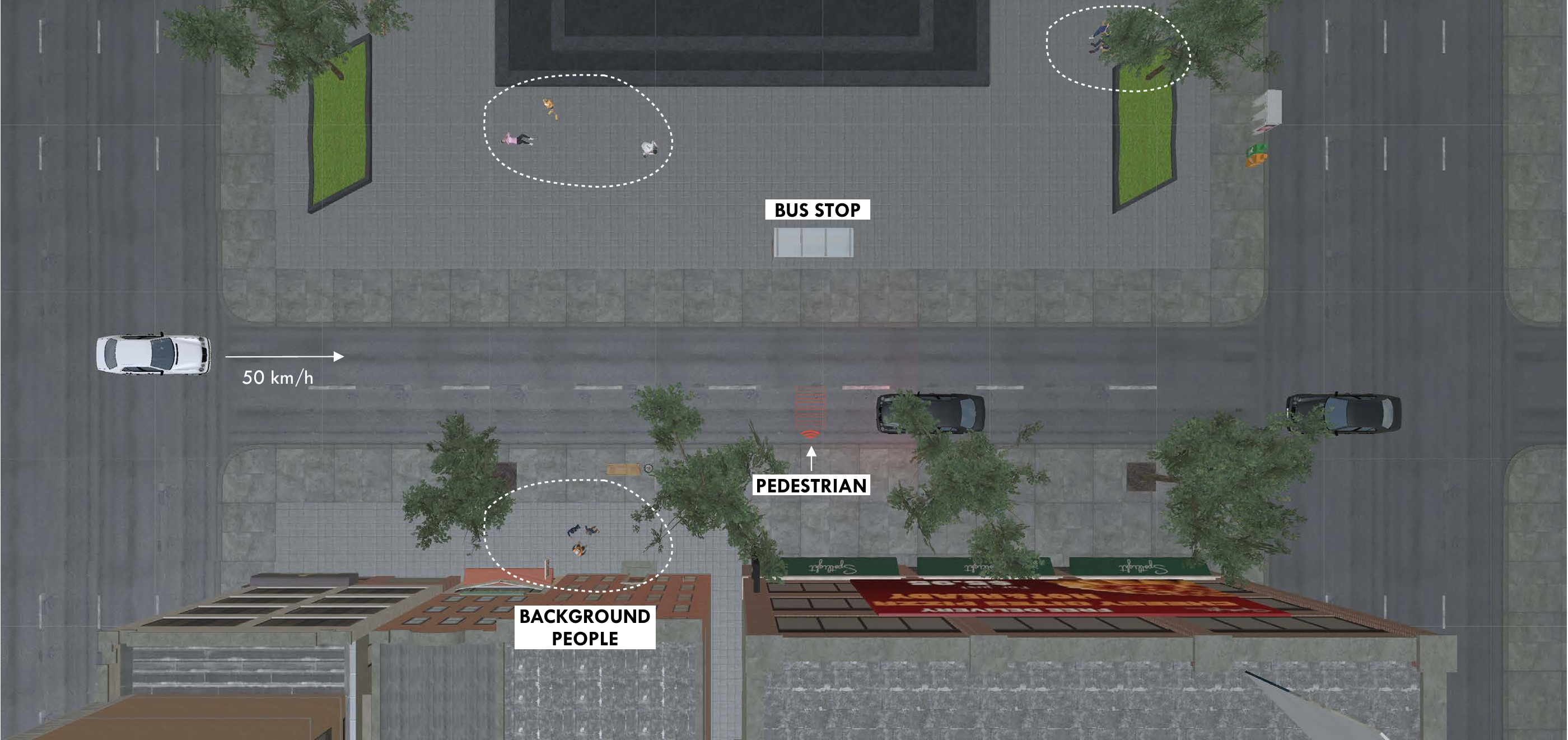}
  \caption{From a bird's-eye view, the simulated environment presents a bus stop opposite the pedestrian's starting point. Dotted lines highlight three groups of background people. The view captures a moment when the AV's eHMI signals to the pedestrian it has stopped, but cautions that the adjacent lane may still be unsafe.}
 \Description{The figure provides a bird's-eye perspective of a simulated environment. At the center stage is a pedestrian's starting point, and directly opposite this is a bus stop. Dotted lines surround three distinct clusters of background people. On the first lane, a black AV has stopped, projecting a red crosswalk sign onto the road. Adjacently, on the second lane, a white manually-driven vehicle is observed advancing at a speed of 50km/h, further accentuating the potential risk for crossing pedestrians.}
\label{fig:birdeyedview}
\end{figure*}

\textit{Hardware:} The Oculus Quest 2, a standalone and untethered VR headset, was employed in this study. Its design allows participants to physically cross a simulated street without being constrained by cables and cords. The headset has a resolution of 1832 x 1920 pixels per eye and an approximate field of view of around 90-95 degrees, depending on factors like the fit of the headset and the user's interpupillary distance. Participants used a Touch controller to navigate from one scenario to the next. Experiments took place in a space with dimensions of 8 m x 5 m (see~\autoref{fig:studysetup}).

\textit{Virtual Environment:} The virtual setting was developed in the Unity 3D game engine\footnote{\url{https://unity.com/}, last accessed February 2024}, using readily available assets from the Unity Asset Store. It depicted an unmarked midblock location on a two-way two-lane urban road, with traffic originating from both directions. Each lane measured 3.5 m in width~\cite{tran2021review}. To provide context for the pedestrian crossing task, a bus stop was situated on the opposite roadside. To enhance the realism of the environment, Mixamo\footnote{\url{https://mixamo.com/}, last accessed February 2024} 3D characters were incorporated to mimic typical human sidewalk activities, such as conversing and exercising. These characters are positioned at a distance to not distract or influence participants' behaviours~\cite{hoggenmueller2021context} (see~\autoref{fig:birdeyedview}). An urban auditory backdrop, featuring bird chirps and traffic noises, was also integrated.

In the simulation, black vehicles symbolised AVs, devoid of a driver, while white vehicles represented manually operated vehicles with a visible driver. The colour distinction was solely to facilitate easier identification, especially in the Baseline condition where AVs are not equipped with an eHMI. Vehicles were spawned from a location outside the participant's line of sight, travelling with a maximum speed of 50 km/h. When 50 m away, they decelerated at a rate of 3.5 $m/s^2$, coming to a full stop at 3 m from a pedestrian, adhering to recommended stopping distances~\cite{qldgov2023}. Emergency braking was not implemented to ensure that any collision events could be directly attributed to the design concepts rather than automated interventions. In each lane, mixed traffic is generated with a sequence of vehicles: AV, Manual, AV, AV, Manual, AV, and then the sequence repeats. The gap between these vehicles is 2.5 - 6.5 seconds. 

\begin{figure*}[htbp]
  \centering
  \includegraphics[width=0.98\linewidth]{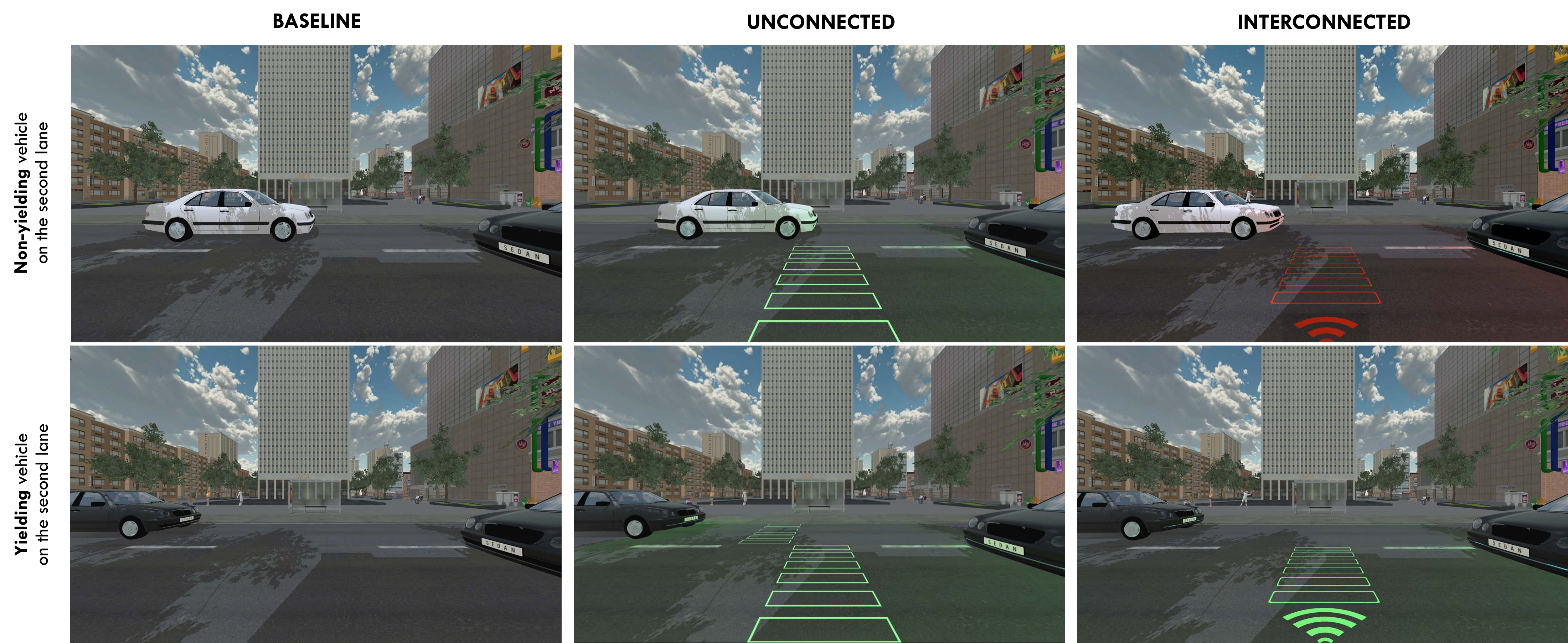}
  \caption{VR implementation of three interface conditions: Baseline without eHMI (left), Unconnected eHMIs (middle), and Interconnected eHMIs (right). The top row illustrates eHMI communication for pedestrians in the presence of a non-yielding vehicle in the second lane, while the bottom row depicts the same with a yielding vehicle.}
  \Description{The figure offers a VR depiction of three distinct interface conditions laid out side-by-side: 'Baseline without eHMI' on the left, 'Unconnected eHMIs' in the middle, and 'Interconnected eHMIs' on the right. The top row illustrates eHMI communication for pedestrians in situations where a vehicle in the second lane isn't yielding. In contrast, the bottom row depicts this communication when the vehicle is yielding.}
  \label{fig:studyconditions}
\end{figure*}

\textit{Design Elements:} To closely mirror real-world implementations of the LED light band, we referenced related field studies~\cite{loew2022goahead} and developed a custom 3D model of the light band. The light band was affixed to the lower front of the vehicle and activated when braking at a distance of 50 m. For the zebra crossing, we used an emissive material to give it a glowing effect.  Additionally, a point light was placed on the AV bumper to enhance the brightness of the projection and indicate the light's origin. In Unity, we employed this approach to achieve the appearance of a laser projection. The Wi-Fi signal displays waves radiating outward. Refer to~\autoref{fig:studyconditions} for the VR implementation of these design elements across interface concepts.

\subsection{Study Procedure} 

Three days prior to the study, all participants received information about the eHMI light band via email. However, participants from the \textit{knowledge} group received additional information about interconnected eHMIs. The material was refined based on insights from a small-scale pilot study involving four people and can be found in Appendix A. On the day of the study, participants signed a consent form and were briefed about the study. Researchers then confirmed that participants had a good understanding of the previously provided material by asking them to explain the concept. This was followed by a familiarisation session in which participants were introduced to both the VR equipment and the virtual environment.

Within this virtual setting, participants learned two distinct types of stationary vehicles. Following this, they had the opportunity to practise crossing the street within the VR environment at least twice, without the presence of vehicles. To contextualise the task, they were instructed to cross the street to reach the bus stop located on the opposite side. 

Before initiating the experiment, participants were reminded to behave as they would under real traffic conditions, meaning they were expected to cross the street only when they deemed it safe to do so. If they found no suitable crossing opportunities, they were advised to inform the researchers to advance to the next scenario. It was clarified that deciding not to cross the street would not impact the outcome of the study in any manner. These instructions were provided to ensure that participants didn't feel pressured to cross the street merely to fulfil the task.

\begin{figure*}[htbp]
  \centering
  \includegraphics[width=1\linewidth]{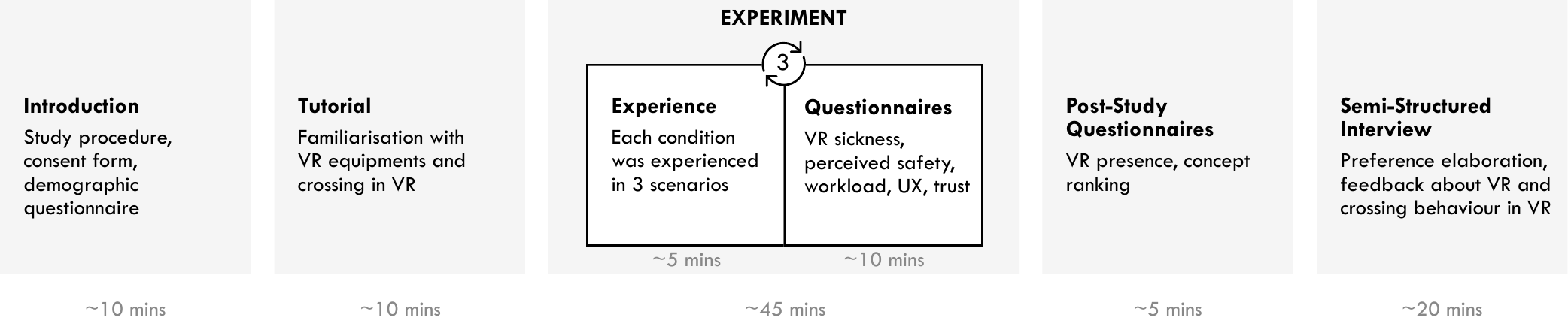}
  \caption{Study procedure and the approximate time for each part of the study.}
  \Description{The figure provides an overview of a study's structure, detailing the duration for each phase. The study starts with an 'Introduction' phase, covering the study procedure, consent form, and a demographic questionnaire, which takes about 10 minutes. This is followed by the 'Tutorial' phase, also lasting 10 minutes, during which participants familiarise themselves with VR equipment and virtual crossing. The 'Experience' phase introduces participants to three conditions, with each condition having three scenarios. Participants spend approximately 5 minutes on the scenarios for each condition. Subsequently, they fill out questionnaires on VR-related subjects for about 10 minutes. The 'Post-Study Questionnaires' phase, which lasts roughly 5 minutes, centres on VR presence and concept evaluations. The study wraps up with a 20-minute 'Semi-Structured Interview', during which participants share their perspectives on the VR experience and virtual crossing.}
  \label{fig:studyprocedure}
\end{figure*}

Participants subsequently underwent three experimental blocks, each comprising three scenarios. Following each block, participants were asked to briefly recall the events and respond to paper questionnaires. Upon completion of all blocks, participants were requested to rank the conditions based on their overall preference and engage in semi-structured interviews. The study lasted between 60 to 90 minutes, depending on the participant's familiarity with VR. An overview of the procedure can be seen in~\autoref{fig:studyprocedure}. 

\subsection{Participants}

We recruited participants from the university's mailing list, social media networks, and through word of mouth. Our sample consisted of 32 individuals, including 20 males and 12 females. Among them were both working professionals and university students, all of whom were fluent in English. To be eligible, participants needed to have normal vision or use glasses to achieve corrected-to-normal vision. Participants with glasses were permitted to wear them with the headset, and all were given ample time to adjust the headset for comfort. Colour blindness was an exclusion criterion. Those with mobility impairments that might hinder movement within a VR environment were also excluded from participation. Each participant was compensated with a \$20 gift card. The study received approval from the university's Human Ethics Committee (ID 2020/779), with which the authors are affiliated. For a detailed breakdown of the demographics and VR experience of the participants, refer to \autoref{tab:participants}.

\begin{table}[htbp]
  \small
  \caption{Demographics and prior experience of participants in two experimental groups.}
  \label{tab:participants}
  \begin{tabular}{lll}
    \toprule
    &\textbf{No Knowledge}&\textbf{Knowledge}\\
    \midrule
    \textbf{N (m/f)} & 16 (10/6) & 16 (10/6) \\
    \textbf{Ages}\\
    18-24 & 4  & 5 \\
    25-34 & 11  & 8  \\
    35-44 & 1 & 2 \\
    55-64 & 0  & 1 \\
    \textbf{VR Experience}\\
    Never & 4 & 1\\
    Less than 5 times & 5 & 8\\
    More than 5 times & 7 & 7\\
    \textbf{AV Experience}\\
    None & 13 & 14\\
    ADAS* & 3 & 2\\
  \bottomrule
  \addlinespace
  \multicolumn{3}{p{6.5cm}}{\textsuperscript{*}Advanced Driver-Assistance Systems (ADAS) features such as as automated braking and lane keeping assistance.}
\end{tabular}
\end{table}

\subsection{Data Collection}

We employed both subjective (questionnaires) and objective (simulation-logged data) methods for data collection.

\textit{Safety}: The Head-Mounted Display (HMD) recorded positional data (x, y, and z coordinates) and logged user events at 60 Hz. This data facilitated the collection of metrics related to safety and crossing behaviour. Key metrics included the number of collisions, crossing initiation time, and crossing duration per lane. These metrics have been utilised in related studies to holistically assess pedestrian safety~\cite{hollander2022take, deb2018investigating}. In the simulation, a collision with the vehicle triggers a fast beep sound, alerting participants to the collision and immediately pausing the scenario, thus minimising potential psychological effects.

\textit{Perceived Safety}: We assessed perceived safety using a 3-item questionnaire with a 5-point Likert scale (based on \citet{hollander2022take}). 

\textit{Workload}: We used NASA Task Load Index~\cite{hart1988development} to assess workload in six dimensions: mental, physical, temporal demand, performance, effort, and frustration. Each of these dimensions is rated on a scale from 0 (very low) to 100 (very high), used to compute an overall workload score. 

\textit{Trust}: Trust in AVs was evaluated using the Trust In Automation questionnaire~\cite{korber2019theoretical}, focusing on three subscales: Reliability/Competence, Understandability/Predictability and Trust in Automation.

\textit{Motion Sickness and Presence:} We used the Misery Scale~\cite{bos2010effect} to monitor simulator sickness, suspending the study if ratings exceeded three. The Igroup Presence Questionnaire{\footnote{\url{https://www.igroup.org/pq/ipq/index.php}, last accessed February 2024}~\cite{schubert2003sense} assessed the degree of participants' immersion in the virtual environment, serving as an indicator of simulation validity. 

\textit{Semi-structured Interviews:} At the end of the study, we asked participants to rank eHMI concepts from 1 (most preferred) to 4 (least preferred) and elaborate on their preferences. We aimed to understand how different eHMI concepts affected their experiences and elements shaping their crossing decisions. Additionally, participants gave feedback on the virtual environment and indicated if their crossing behaviour in the VR setting mirrored their actions in the real world.

\subsection{Data Analysis}

\subsubsection{Quantitative Analysis} 
We employed IBM SPSS Statistics Version 29.0.1.0 for all statistical analyses. Our initial step was to assess the internal reliability of the questionnaires~\cite{devellis1991scale}. The \textit{Perceived Safety} scale demonstrated good reliability ($\alpha = 0.803$), and the \textit{Workload} scale was similarly robust ($\alpha = 0.885$). For the \textit{Trust in Automation} questionnaire subscales: \textit{Reliability/Competence} achieved good reliability ($\alpha = 0.810$), \textit{Understandability/Predictability} was deemed acceptable ($\alpha = 0.727$), and \textit{Trust in Automation} was also good ($\alpha = 0.888$). Regarding the \textit{Presence} questionnaire, the overall scale indicated acceptable reliability ($\alpha = 0.754$).

We then moved to preliminary descriptive analyses and plotting of the data. Given the hierarchical structure of our study design—where scenarios are nested within each condition—and to account for repeated measures, we applied the Linear Mixed Model (LMM). Within the LMM, we set Group (knowledge vs. no knowledge) and Condition (Baseline, Unconnected, Interconnected) as fixed effects, and participants were treated as a random effect to accommodate inter-individual differences.
Initially, we considered including a nested random effect for the scenarios. However, upon further evaluation, we recognised that incorporating nested random effects significantly increased the model's complexity. To minimise the risk of overfitting, we decided to exclude the nested random effects. For pairwise comparisons, we applied Bonferroni adjustments to mitigate the risk of Type I errors from multiple tests. Adjusted p-values are reported and assessed for significance.

\subsubsection{Qualitative Analysis} 
We employed a deductive thematic analysis process to analyse the data that pertained to our research questions. Two coders were involved in the analysis: Coder 1 who designed the study and conducted the interviews, and Coder 2, who was not part of these processes. Coder 1 coded all the interviews from the participant group with prior knowledge about interconnected eHMIs, while Coder 2 coded all the interviews from the other group. This division of coding responsibilities allowed for a focused analysis of each group's responses. Moreover, each coder selected a subset of six interviews from their respective groups. These subsets, chosen for their representation of the range of responses, underwent double coding to ensure consistency and reliability in the analysis. 

\section{Results}

This section combines both quantitative and qualitative insights for a holistic presentation of the results. We begin with the effects of conditions on different measures, followed by analyses of group effects, interaction effects, and random effects. For ease of reference of qualitative results, participants from the \textit{knowledge} group are highlighted in violet and referenced with the subscript notation \textcolor{violet}{$n_k$}. Those from the \textit{no knowledge} group are highlighted in teal and referenced as \textcolor{teal}{$n_{nk}$}. An overview of qualitative observations, contrasting Unconnected and Interconnected eHMIs, can be found in~\autoref{tab:ehmi_feedback}.

\subsection{Safety}

\subsubsection{Collisions}

In 288 trials, there were 7 collisions in the Baseline condition, 10 in the Unconnected, and 12 in the Interconnected (see \autoref{tab:combined}.
Collisions occurred in the second lane and across both groups: no knowledge (16) and knowledge (13). The majority of participants had either no collisions (9) or just one collision over nine trials (17). This pattern suggests that most participants adapted their approach to reduce further collisions in subsequent trials. Nonetheless, a few exceptions were noted: 6 participants encountered two collisions.

Post-collision interviews revealed the factors contributing to these collisions are:

\begin{itemize}
    \item Misjudgment of the distance or time needed to safely cross the street, which was influenced by VR simulation limitations (13 collisions, 45\% of all collisions).
    \item Expectations or assumptions about vehicle behaviour, particularly with AVs, with participants believing that these vehicles would always stop or at least slow down for them (10, 35\%).
    \item Behavioural tendencies, such as avoiding rushing or a willingness to take risks (3, 10\%).
    \item Misinterpretation of eHMI communication (3, 10\%).
\end{itemize}

Delving deeper into the eHMI misinterpretation factor, three participants highlighted a recurring misconception of unconnected eHMIs. They misinterpreted the extent of safety the green light indicated, believing it signified a safe passage for the entire street when it only applied to one lane. While P10 \textit{`did not check the other direction'}, P16 mentioned \textit{`[seeing] the green and went across the entire street'}. Both P10 and P16 belonged to the \textit{no knowledge} group. The third participant, P24, who was in the \textit{knowledge} group, expressed that the misinterpretation was influenced by prior knowledge about interconnected eHMIs and thought the vehicles were also connected to this unconnected eHMI concept. 

Given that collision data per condition are count values ranging from 0 to 1 (with a single exception of 2) for each condition, we employed a Generalised Linear Model to estimate the probability of a collision across different interface conditions. To assess collision occurrences, we introduced the binary variable HasCollisions, which indicates whether a collision (coded as `Yes') or no collision (coded as `No') was observed for each data point. The analysis revealed that the p-value for the main effect of Condition was 0.384, suggesting that, based on the current data and model, there was no statistically significant relationship between Condition and HasCollisions.

However, despite achieving an overall classification accuracy of 69.8\%, it's important to note that the model consistently failed to predict `Yes' for HasCollisions, suggesting that the model may be biased towards the majority class (`No'). This bias can primarily be attributed to the significant imbalance in the dataset, with `No' outcomes far outnumbering `Yes' outcomes.

\begin{table*}[htbp]
\centering
\small
\caption{Descriptive statistics for Collisions, Crossing Initiation Time (CIT), Time on First Lane, and Time on Second Lane}
\label{tab:combined}
\begin{tabular}{llccccccc}
\toprule
\multirow{2}{*}{\textbf{Condition}} & \multicolumn{4}{c}{\textbf{Collisions on Second Lane}} & \textbf{CIT} & \textbf{Time on First Lane} & \textbf{Time on Second Lane} \\
\cmidrule{2-5}
& \textbf{Total} & AV yields & Manual yields & No yield & \textbf{Mean}/SD & \textbf{Mean}/SD & \textbf{Mean}/SD \\
\midrule
Baseline & \textbf{7 (24\%)} & 2 & 0 & 5 & \textbf{12.76}/1.03 & \textbf{5.51}/5.10 & \textbf{4.01}/2.065 \\
Unconnected & \textbf{10 (35\%)} & 2 & 3 & 5 & \textbf{10.50}/1.02 & \textbf{5.72}/5.33 & \textbf{3.72}/1.595 \\
Interconnected & \textbf{12 (41\%)} & 1 & 5 & 6 & \textbf{12.17}/1.14 & \textbf{5.12}/4.71 & \textbf{3.90}/1.612 \\
\bottomrule
\end{tabular}
\end{table*}

\subsubsection{Crossing Initiation Time (CIT)} This is defined as the time taken by the participant to start crossing from the moment an AV starts to slow down on the nearest lane. When participants began crossing before a vehicle started yielding, we assigned a negative crossing initiation time. This value represents the seconds before the vehicle's yield onset. Out of 288 trials, there were 5 scenarios with negative CITs (1.74\%).

There was a significant main effect of Condition on CIT, F(2, 113.29)~=~3.16, p~=~.046. Post hoc tests revealed that CIT was significantly longer in the Baseline condition compared to the Unconnected condition (Mdiff~=~2.26, SE~=~0.93, p~=~.048). No significant differences in CIT were observed between Baseline and Interconnected or between Unconnected and Interconnected.

\subsubsection{Time on First Lane} Out of 288 trials, there were 3 (1\%) in which participants did not step onto the first lane, choosing instead not to cross. These specific scenarios occurred when the vehicle did not stop on the second lane. Additionally, we identified 32 outliers (11\%) using boxplot visualisations. To mitigate the influence of these outliers, we used a winsorising method, replacing extreme values with the 5th and 95th percentiles based on the data distribution. Descriptive analysis of the adjusted dataset highlighted only minor differences in mean times across conditions, and the variability within each condition was notably greater than any differences between them. Given this observation, further in-depth analysis on this specific metric was deemed unnecessary.

\subsubsection{Time on Second Lane} Of 288 trials, 37 (12.85\%) lacked data for the second lane due to non-crossing (7), collisions (29), or tracking error (1). For the trials that rendered data, the descriptive statistics suggests minor differences in mean times across these conditions, with variability within each condition being larger than differences between them. 

\subsection{Perceived Safety}

\subsubsection{Questionnaire}
There was a significant main effect of Condition on perceived safety, F(2, 42.53)~=~6.81, p~=~.003. Post hoc tests revealed that perceived safety was significantly lower in the Baseline condition compared to both Unconnected (Mdiff~=~-0.32, SE~=~0.12, p~=~.032) and Interconnected (Mdiff~=~-0.39, SE~=~0.11, p~=~.003) conditions. However, the Unconnected and Interconnected conditions did not differ significantly (see~\autoref{fig:condition_boxplot}).

\subsubsection{Qualitative Insights}

When no eHMI was present, several participants felt safe as vehicles stopped for them at non-priority pedestrian crossings \textcolor{teal}{\textbf{({$n_{nk}$}=2)}}. Some mentioned feeling comfortable because having no eHMI was closer to their everyday experiences, and they simply treated the AVs as regular vehicles (\textcolor{teal}{\textbf{{$n_{nk}$}=2}}, \textcolor{violet}{\textbf{{$n_k$}=2}}). The majority, however, felt unsafe due to the lack of vehicular signals. They were uncertain if they had been \textit{`seen'} and feared the vehicle might resume driving (\textcolor{teal}{\textbf{{$n_{nk}$}=8}}, \textcolor{violet}{\textbf{{$n_k$}=10}}).

Participants perceived the green crosswalk of unconnected eHMIs as effectively signaling a safe-to-cross message, largely due to the widespread association of green with safety and positive feelings such as \textit{`safe'}, \textit{`confident'}, and \textit{`comfortable'} (\textcolor{teal}{\textbf{{$n_{nk}$}=22}}, \textcolor{violet}{\textbf{{$n_k$}=10}}). However, after crossing the first lane, there was evident uncertainty, particularly concerning the intentions of vehicles in the second lane \textcolor{teal}{\textbf{({$n_{nk}$}=6}, \textcolor{violet}{\textbf{{$n_k$}=7)}}}. P1 posed the question, \textit{`it's safe to cross half the road, but what about the other half?'} There were concerns about endangering children and individuals engrossed in their phones \textcolor{violet}{\textbf{({$n_k$}=2)}}. In contrast, five participants recognised that the signal pertained only to the first lane, basing this interpretation on the observation that the crosswalk covered only half the street \textcolor{violet}{\textbf{({$n_k$}=5)}}.

The red crosswalk of the interconnected eHMIs triggered cautionary behaviours among pedestrians, eliciting a range of responses: from outright hesitation and reluctance to cross \textcolor{teal}{\textbf{({$n_{nk}$}=6)}}, to increased attentiveness to surrounding traffic conditions \textcolor{teal}{\textbf{({$n_{nk}$}=2)}}. For instance, P8 remarked, \textit{`When the signal is red, it's definitely a no for me to even think of crossing at all. I didn't even cross half of the road.'} Meanwhile, P2 perceived the red as a sign of vigilance, noting, \textit{[It] was more of an alert, like ``stay alert''.'} However, many relied on the presence of the crosswalk and the fact that the vehicle had stopped to proceed as usual \textcolor{teal}{\textbf{({$n_{nk}$}=7)}}. P15 commented, \textit{It resembled a zebra crossing, so I crossed even though the light was red. I didn't think much about it.'} Among participants who were briefed about the concept, more considered the red crosswalk a reminder to be vigilant in the crossing situation \textcolor{violet}{\textbf{({$n_k$}=10)}}. P27 thought \textit{`It is good to have AVs assisting humans in decision-making while crossing the road. However, we shouldn't encourage people to become fully dependent on AVs, regardless of their reliability.'} Many participants felt a heightened sense of safety, either because the additional information about the broader traffic situation served as an \textit{`extra layer of safety'} \textcolor{teal}{\textbf{({$n_{nk}$}=6)}} or because of the interconnectivity \textcolor{violet}{\textbf{({$n_k$}=6)}}. As several participants pointed out, \textit{`I crossed because I understood that the cars could connect with each other and form a cohesive system,'} said P21.

\begin{figure*}[htbp]
  \centering
  \includegraphics[width=0.98\linewidth]{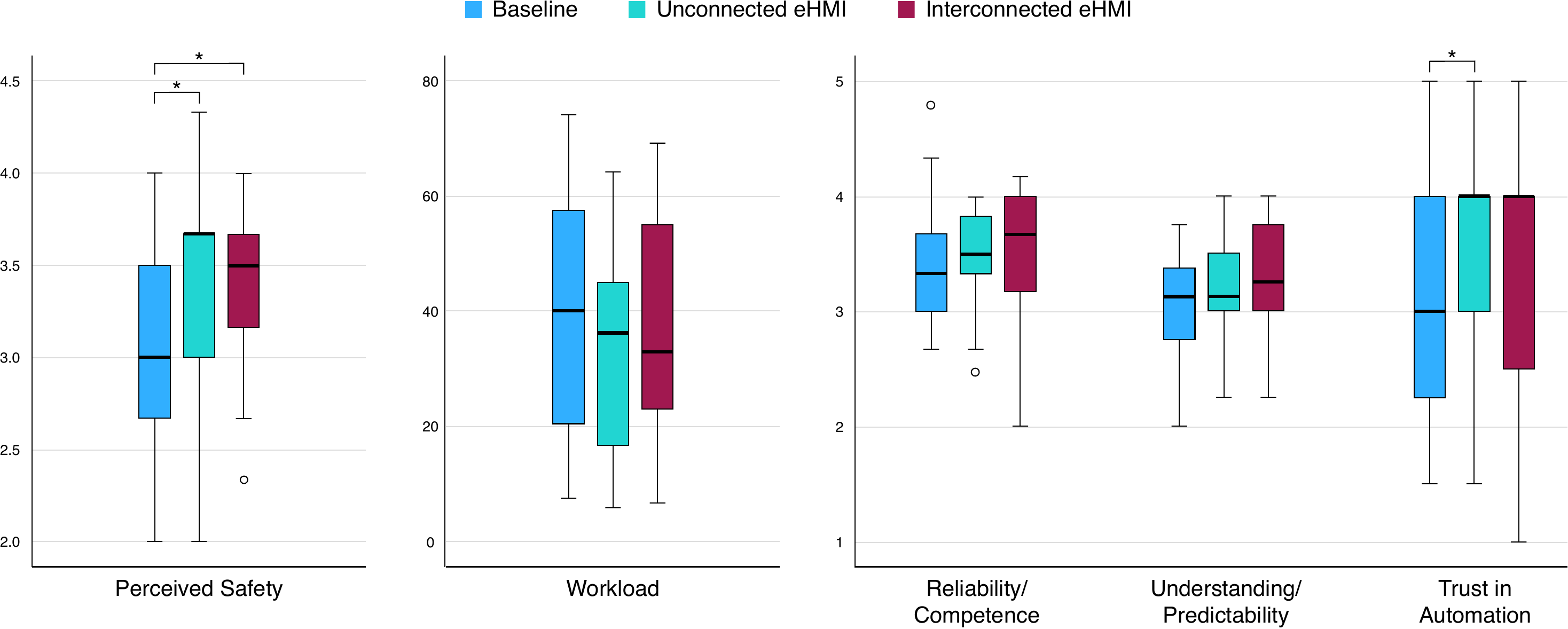}
  \caption{Box plots of Perceived Safety, Workload, and Trust subscales across different conditions.  * indicates \( p < .05 \)}
  \Description{The figure presents three box plots arranged side-by-side, each representing different subscales measured across various conditions. The leftmost plot illustrates the distribution of 'Perceived Safety' scores, the middle one displays 'Workload' scores, and the rightmost plot depicts 'Trust' scores. Each plot shows typical statistical markers such as median, quartiles, and potential outliers. Notably, data points or conditions highlighted with an asterisk (*) indicate statistical significance, having a p-value less than 0.05.}
  \label{fig:condition_boxplot}
\end{figure*}

\subsection{Workload}

\subsubsection{Questionnaire}

Descriptively, the Baseline condition recorded the highest mean (M~=~40.37, SD~=~20.51), followed by the Interconnected condition (M~=~37.63, SD~=~18.02), and the Unconnected condition (M~=~33.13, SD~=~16.38). Though there were observable differences in mean scores across conditions, there was no significant main effect, F(2, 27.99)~=~3.21, p~=~.056 (see~\autoref{fig:condition_boxplot}).

\subsubsection{Qualitative Insights} Participants were found to invest considerable cognitive resources attempting to decipher the vehicle's intentions, eHMI signals, or the overall traffic situation.

In the Baseline condition without an eHMI, the only signal available to pedestrians was the vehicle's movement. This condition introduced ambiguity as pedestrians often questioned the vehicle's intentions: Would it stop, continue driving, or resume driving shortly? (\textcolor{teal}{\textbf{{$n_{nk}$}=9}}, \textcolor{violet}{\textbf{{$n_k$}=5}}). Even when vehicles stopped, it was unclear to pedestrians whether the stop was meant for their crossing or due to other reasons, such as a technical error (\textcolor{teal}{\textbf{{$n_{nk}$}=3}}, \textcolor{violet}{\textbf{{$n_k$}=8}}). However, for several other participants, the absence of AV communication simply meant waiting a bit longer to ensure the vehicles remained stationary (\textcolor{teal}{\textbf{{$n_{nk}$}=5}}, \textcolor{violet}{\textbf{{$n_k$}=7}}).

In regards to the unconnected eHMIs, they were perceived as somewhat inferior in terms of the information they provided \textcolor{violet}{\textbf{({$n_k$}=3)}}. As P29 pointed out, \textit{`While the green colour is like, ``Oh, yeah. Everything is good," it provides less information, which can be risky.'} Nevertheless, the green crossing signal was widely acknowledged for its clarity and ease of comprehension. This sentiment was universally shared among participants, irrespective of their prior knowledge~\textcolor{teal}{\textbf{({$n_{nk}$}=14}}, \textcolor{violet}{\textbf{{$n_k$}=9)}}.

With the interconnected eHMIs, several themes pertaining to cognitive load were identified:

\begin{itemize}

    \item \textbf{Red/Green Encoding}: The red/green encoding was met with varied reactions. While the red colour traditionally signifies  `do not cross', a crosswalk inherently suggests safe passage. This dichotomy resulted in confusion among pedestrians regarding the appropriate action \textcolor{teal}{\textbf{({$n_{nk}$}=5)}}. Furthermore, with the red crosswalk being projected from a stopped AV, participants \textcolor{teal}{\textbf{({$n_{nk}$}=4)}} were left pondering, \textit{`Why should I wait?'} or getting frustrated \textit{`you see the road is clear, nothing obstructs you. So you may feel a bit more irritated, like it's turning red, and you're not allowed to go'}. Interestingly, even those familiar with the design concept were not spared from this confusion \textcolor{violet}{\textbf{({$n_k$}=8)}}. P21 attributed this to \textit{`everyday experiences,'} while P30 described it as \textit{`a deeply ingrained response'} to the colour red. On the other hand, certain participants found the red/green encoding to be straightforward \textcolor{teal}{\textbf{({$n_{nk}$}=5)}}. As P1 noted, \textit{`I like the concept of the red light indicating that it isn't safe to proceed. Then, when it turned green, the message was clear: ``You can now walk''.'} 
    
    \item \textbf{Wi-Fi Signal}: Feedback from respondents revealed a consistent sense of uncertainty regarding the meaning of the Wi-Fi signal \textcolor{teal}{\textbf{({$n_{nk}$}=11)}}. Many mentioned they had no idea about its significance \textcolor{teal}{\textbf{({$n_{nk}$}=9)}}. Additionally, the Wi-Fi symbol sparked a diverse range of interpretations \textcolor{teal}{\textbf{({$n_{nk}$}=12)}}. These interpretations varied, with participants suggesting it might represent autonomous technology, act as a sensing mechanism, or indicate vehicle-to-vehicle communication. A smaller subset felt that the spreading-out animation of the symbol suggested an invitation to cross \textcolor{teal}{\textbf{({$n_{nk}$}=5)}}.

    \item \textbf{Additional Layer of Information}: From an information perspective, the interconnected eHMI incorporated information about the safety of both lanes—a nuance that participants from both groups recognised and valued \textcolor{teal}{\textbf{({$n_{nk}$}=6}}, \textcolor{violet}{\textbf{{$n_k$}=6)}}. However, some reservations were also noted \textcolor{teal}{\textbf{({$n_{nk}$}=2}}, \textcolor{violet}{\textbf{{$n_k$}=3)}}. P2 felt the information might be \textit{`too much for non-tech and late adopters,'} while P25 expressed concerns that it could slow down the crossing process. 
\end{itemize}

\subsection{Trust}

\subsubsection{Questionnaire}

There was no significant main effect of Condition on the Reliability/Competence subscale, F(2, 34.36)~=~.90, p~=~.417. The main effect of Condition on the Understandability/Predictability subscale was not statistically significant, F(2, 35.65) = 2.63, p~=~.086.

There was a significant main effect of Condition on Trust In Automation subscale, F(2, 38.44) = 3.53, p = .039. Post hoc tests revealed that the trust score was significantly lower in Baseline compared to Unconnected (Mdiff~=~-0.52, SE~=~.20, p~=~.036). However, there was no significant difference in trust scores between Baseline and Interconnected (Mdiff~=~-0.38, SE~=~0.21, p~=~.243), or between Unconnected and Interconnected (Mdiff~=~0.14, SE~=~0.19, p~=~1.000).

\subsubsection{Qualitative Insights}
Comments about the Baseline condition reflected a lack of confidence in the vehicle's behaviour. As P6 stated, \textit{`if I start crossing and the car, for some reason, thinks there's no one in front, it might start speeding up and could run over me'} (\textcolor{teal}{\textbf{{$n_{nk}$}=7}}, \textcolor{violet}{\textbf{{$n_k$}=7}}). As a result, there was a prevalent desire for clear indications from vehicles to enhance predictability and reliability (\textcolor{teal}{\textbf{{$n_{nk}$}=6}}, \textcolor{violet}{\textbf{{$n_k$}=7}}).

The unconnected eHMIs were perceived as trustworthy due to its simplicity and consistency (\textcolor{teal}{\textbf{{$n_k$}=7}}, \textcolor{violet}{\textbf{{$n_k$}=4}}). Participants remarked, \textit{`the green one felt more reliable because it didn't change'} (P2). However, some described the concept as imparting a false sense of safety (\textcolor{teal}{\textbf{{$n_k$}=2}}, \textcolor{violet}{\textbf{{$n_k$}=4}}), labeling it as \textit{`misleading'} (P18) and \textit{`deceptive'} (P29). P18 offered a perspective, noting, \textit{`the projected green crossing gave me slightly false information [...] it seemed to indicate that I could cross more than it actually meant.'}

The interconnected eHMIs were commended for providing reliable information about traffic situations (\textcolor{teal}{\textbf{{$n_{nk}$}=8}}, \textcolor{violet}{\textbf{{$n_k$}=10}}). As P11 stated, \textit{`the crossing indicates whether it's safe to cross on just one lane or both lanes of the road. It provides more reliable information for me to make a decision.'} The adaptability of the interface played a role in building trust, a sentiment echoed by P5, who stated, \textit{`the fact that the interface changed based on the vehicle's awareness of other vehicles made me trust it more because it felt like it was looking out for my safety.'} Nevertheless, skepticism arose when participants could not understand the purpose of the Wi-Fi symbol or found it inconsistent with their expectations (\textcolor{teal}{\textbf{{$n_{nk}$}=2}}): \textit{`I assumed the Wi-Fi symbol indicated a communication mode between the cars, but seeing a car move ahead without yielding made me question its reliability'} (P12). Similarly, in scenarios where the AV didn't yield on the second lane, some participants expressed trust concerns about connectivity (\textcolor{violet}{\textbf{{$n_k$}=3}}). They felt that, \textit{`when one car has already stopped to give way to a pedestrian, the other car should take that into consideration and stop as well'} (P25). Doubts arose among some participants when communication came solely from one AV (\textcolor{violet}{\textbf{{$n_k$}=3}}), with comments like, \textit{`when all the information is coming from one car, I wouldn't have that confidence'} (P24).

\subsection{Group Effects and Interaction Effects}

The Group Effect assesses the differences in responses between the two groups: the \textit{knowledge} and the \textit{no knowledge} group. From our results, there were no significant differences between these groups in terms of their workload and perceived safety. Regarding trust subscales, Trust In Automation and Reliability/Competence also did not show significant differences between the two groups. However, the Understandability/Predictability subscale presented a notable distinction. The \textit{knowledge} group registered lower scores compared to the \textit{no knowledge} group, as indicated by a value of Mdiff~=~-0.24, SE~=~0.10, p~=~.024. This suggests that the \textit{knowledge} group perceived the system as less predictable or understandable than the \textit{no knowledge} group did.

The Interaction Effect examines how the difference between these two groups changes based on specific interface conditions. The data suggests that the \textit{knowledge} group appears to perceive the Interconnected eHMI more favourably on measures such as perceived safety, workload, and trust subscales when compared to the Unconnected and Baseline conditions (see~\autoref{fig:interaction_lines})
Nonetheless, the interaction effects were not statistically significant (see \autoref{tab:summarystats}). This suggests that the impact of the different interface conditions on these measures remains uniform regardless of whether participants had prior knowledge of interconnected eHMIs.

\begin{figure*}[htbp]
  \centering
  \includegraphics[width=0.98\linewidth]{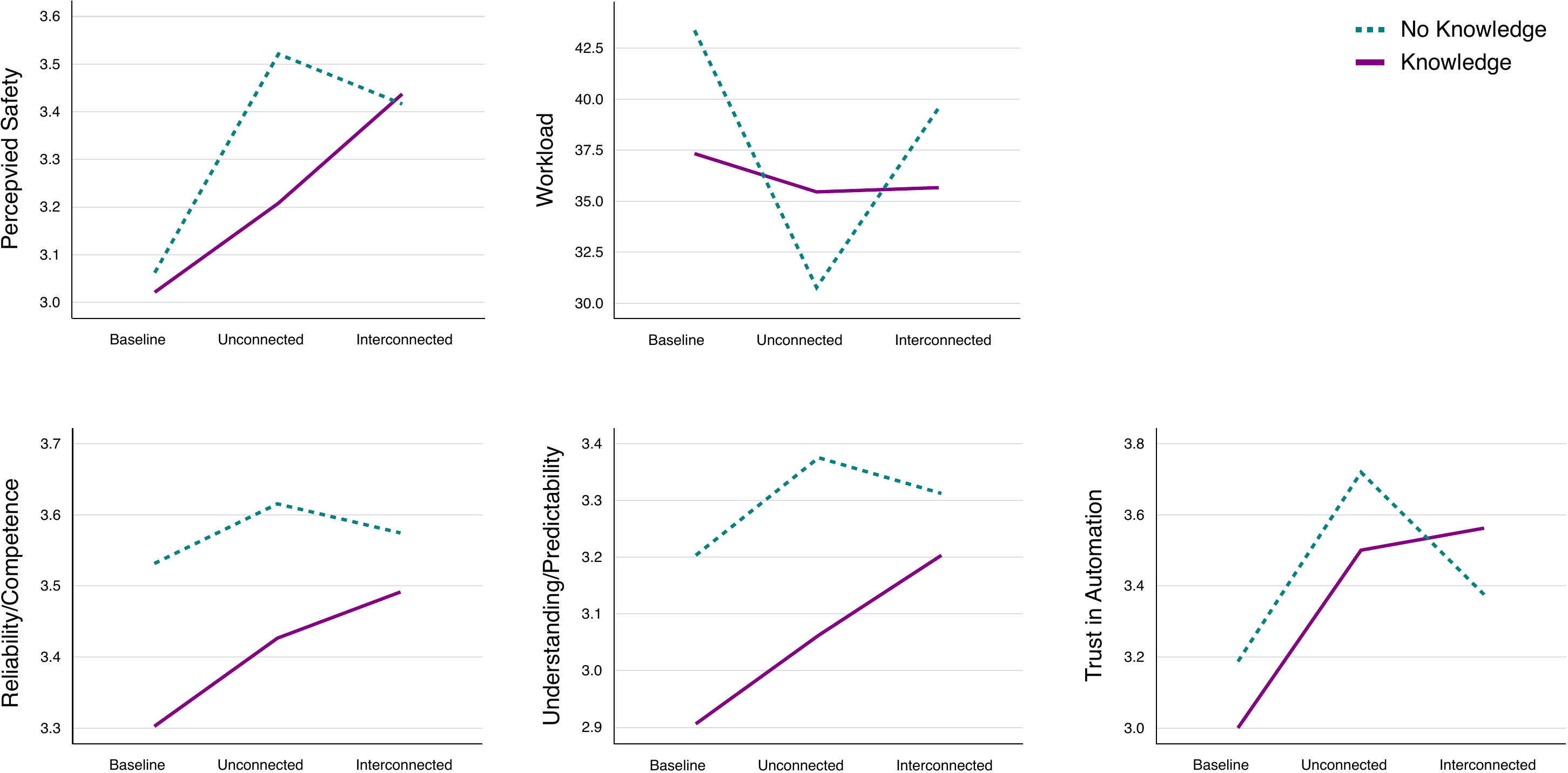}
  \caption{Interaction between Group and Condition across multiple measures: Perceived Safety, Workload, and three Trust subscales. Each line represents the estimated marginal means for each measure, highlighting differential responses between the groups across the different interface conditions.}
  \Description{The figure illustrates the interaction between 'Group' and 'Condition' for several measures: 'Perceived Safety', 'Workload', and three 'Trust' subscales. Lines in the visual represent the estimated marginal means for each measure. These lines provide a visual representation of the responses of the two groups (Knowledge and No Knowledge) to the interface conditions (Baseline, Unconnected, Interconnected).}
  \label{fig:interaction_lines}
\end{figure*}

\begin{table*}[htbp]
\centering
\small
\caption{Summary of Main Effects and Interaction Effects}
\label{tab:summarystats}
\begin{tabular}{llll}
\toprule
\textbf{Dependent Variable} & \parbox[t]{3.35cm}{\textbf{Main Effect} \\ \textbf{(Condition)}} & \parbox[t]{3.35cm}{\textbf{Main Effect} \\ \textbf{(Group)}} & \parbox[t]{3.35cm}{\textbf{Interaction Effect} \\ \textbf{(Group $\times$ Condition)}} \\
\midrule
Perceived Safety & .003* & .383 & .305 \\
Workload & .056 & .748 & .162 \\
Trust - Reliability/Competence & .417 & .227 & .772 \\
Trust - Understandability/Predictability & .086 & .024* & .562 \\
Trust - TIA & .039* & .803 & .518 \\
\bottomrule
\addlinespace
\multicolumn{4}{l}{\textsuperscript{*} indicates \( p < .05 \).}
\end{tabular}
\end{table*}

\subsection{Random Effects}

We observed variability in workload, perceived safety, and trust subscales across different conditions, as presented in Table \ref{tab:random_effects_summary}. The data reveals two primary insights:

\begin{itemize}
    \item Individual Differences: The p (Variance) column indicate the statistical significance of individual differences on Workload, Reliability/Competence, and TIA. The data suggests that the attributes and experiences of participants have a notable influence on these measures across the conditions.
    \item Condition-specific Variations: The p (Condition) column presents the significance levels for each measure under the distinct conditions of Baseline, Unconnected, and Interconnected. 
\end{itemize}

\begin{table*}[htbp]
\centering
\small
\caption{Summary of Variability in Perceived Safety, Workload, and Trust subscales across different conditions}
\label{tab:random_effects_summary}
\begin{tabular}{lllllll}
\toprule
\textbf{Measure} & \textbf{Variance} & \textbf{p (Variance)} & \textbf{Baseline} & \textbf{Unconnected} & \textbf{Interconnected} & \textbf{p (Condition)} \\
\midrule
Perceived Safety & 0.06 & 0.084 & 0.23 & 0.24 & 0.14 & .001/.001/.003 \\
Workload & 171.80 & .006* & 187.53 & 92.43 & 279.05 & .002/.042/.002 \\
Trust - Reliability/Competence & 0.10 & .010* & 0.15 & 0.09 & 0.19 & .004/.013/.002 \\
Trust - Understandability/Predictability & 0.03 & .146 & 0.13 & 0.10 & 0.21 & .002/.004/<.001 \\
Trust - TIA & 0.47 & .008* & 0.76 & 0.46 & 0.65 & .002/.008/.003 \\
\bottomrule
\addlinespace
\multicolumn{7}{p{14.5cm}}{Significance levels for each condition are presented in the last column in the order of Baseline, Unconnected, and Interconnected. \textsuperscript{*} indicates \( p < .05 \)}
\end{tabular}
\end{table*}

\subsection{Motion Sickness and Presence}

The Misery Scale showed that participants experienced either no problems or only vague symptoms of motion sickness. The average Presence score for our 32 participants was 0.43 (SD~=~0.50), with scores ranging from a low of -0.71 to a high of 1.36, suggesting a moderate sense of presence.

\begin{table*}[htbp]
    \small
    \centering
    \caption{Summary of participant feedback on Unconnected and Interconnected eHMI designs by theme. The length of the coloured bars represents the number of participants providing specific feedback, while a dash (-) indicates no feedback was provided.}
    \label{tab:ehmi_feedback}
    \begin{tabular}{lllp{0.1cm}lp{0.1cm}l}
    \hline
    \textbf{Theme} & \textbf{eHMI Type} & \textbf{Feedback} & \multicolumn{2}{l}{\textbf{No Knowledge}} & \multicolumn{2}{l}{\textbf{Knowledge}} \\
    \hline
    \rowcolor{gray!10}
    \multicolumn{7}{l}{\textbf{Safety}} \\
    \hline
    & Unconnected & Green colour induced safe and positive feelings & \scriptsize{22} & \tealblock{22} & \scriptsize{10} & \violetblock{10} \\
    & & Signaled only first lane safety & \scriptsize{6} & \tealblock{6} & \scriptsize{7} & \violetblock{7} \\
    & Interconnected & Red colour triggered cautionary behaviours & \scriptsize{8} & \tealblock{8} & \scriptsize{10} & \violetblock{10} \\
    & & Heightened sense of safety & \scriptsize{6} & \tealblock{6} & \scriptsize{6} & \violetblock{6} \\
    \hline
    \rowcolor{gray!10}
    \multicolumn{7}{l}{\textbf{Cognitive Load}} \\
    \hline
    & Unconnected & Clear and easy to understand & \scriptsize{14} & \tealblock{14} & \scriptsize{9} & \violetblock{9} \\
    & & No information about the second lane & - & - & \scriptsize{3} & \violetblock{3} \\
    & Interconnected & Additional layer of information was appreciated & \scriptsize{6} & \tealblock{6} & \scriptsize{8} & \violetblock{8} \\
    && Red colour caused confusion/hesitation & \scriptsize{9} & \tealblock{9} & \scriptsize{8} & \violetblock{8} \\
    & & Wi-Fi signal caused confusion/uncertainty & \scriptsize{11} & \tealblock{11} & - & - \\
    \hline
    \rowcolor{gray!10}
    \multicolumn{7}{l}{\textbf{Trust}} \\
    \hline
    & Unconnected & Simplicity and consistency promoted trust & \scriptsize{7} & \tealblock{7} & \scriptsize{4} & \violetblock{4} \\
    & & False sense of safety & \scriptsize{2} & \tealblock{2} & \scriptsize{4} & \violetblock{4} \\
    & Interconnected & Reliable information about traffic situations & \scriptsize{8} & \tealblock{8} & \scriptsize{10} & \violetblock{10} \\
    & & Skepticism from the Wi-Fi symbol & \scriptsize{2} & \tealblock{2} & - & - \\
    & & Concerns when AV didn't yield / sole communicator & - & - & \scriptsize{6} & \violetblock{6} \\
    \hline
    \end{tabular}
\end{table*}

\section{Discussion}

\subsection{Effect of Interconnected eHMIs (RQ1)} 

\subsubsection{Measured Safety}

Even though the interconnected eHMI was designed to ensure clearer and integrated communication, it recorded the highest number of collisions. From the qualitative analysis of collision data, none of these collisions resulted from misinterpretation of the eHMI signals. Rather, the collisions were primarily influenced by pedestrians' misjudgement of the distance or time required to cross and their assumptions about vehicle behaviour in the second lane. 

The unconnected eHMI, despite having similar collision numbers, raised concerns about signal misinterpretation. The two collisions in the \textit{no knowledge} group offer empirical evidence of a potential issue extensively discussed in existing literature~\cite{tran2023scoping}. Pedestrians might over-rely on instructional eHMIs for crossing guidance, imparting a false sense of safety and causing them to overlook other crucial environmental cues. Indeed, participants spent the least time on the sidewalk and the longest time on the first lane, suggesting they stepped onto the road as soon as they observed the green crosswalk. \citet{hollander2022take} found a comparable outcome when investigating a similar projection-based concept, and the authors argued that spending more time on the road might increase the risk of pedestrians getting hit, as well as disrupting traffic flow. 

The Baseline condition reported fewer collisions compared to the other conditions. Pedestrians appeared to adopt a more cautious approach, as seen by their longest waiting times on the sidewalk. This behaviour aligns with the mixed traffic study by~\citet{mahadevan2019av}, where 11 of 12 participants explicitly mentioned that the absence of vehicle interfaces made them more vigilant while crossing. A complex traffic scenario, such as the one investigated here, implies that the addition of vehicles of different automation levels and the intricacies of navigating multi-lane crossings can introduce variables that might not have been accounted for in simpler studies. Such scenarios might render eHMIs less effective or even counterproductive. One potential reason is that in the second lane where traffic included both vehicles with and without eHMIs, pedestrians had to discern the potential actions of vehicles based on the eHMI signals or the lack thereof. In the absence of eHMI, as in the Baseline condition, decisions were primarily based on vehicle movement, distance, and speed.

From this, it can be inferred that while interconnected eHMIs reduce issues of signal misinterpretation found in unconnected eHMIs, there are other aspects of their design or implementation that might be affecting their potential to enhance safety. 

\subsubsection{Perceived Safety}

Contrary to actual safety measures, both interconnected and unconnected eHMIs statistically elicited a significantly higher safety perception among participants compared to the baseline condition. Consistent with our observations, broader research also indicates that an enhancement in perceived safety due to AV interfaces is a recurrent finding, irrespective of the research methodology employed. This has been demonstrated in VR studies~\cite{bockle2017sav2p, deb2018investigating, declercq2019external} as well as Wizard-of-Oz-based field experiments~\cite{habibovic2018communicating}. These studies gauge perceived safety using varied tools, from questionnaires~\cite{bockle2017sav2p, habibovic2018communicating, deb2018investigating} to handheld remotes~\cite{declercq2019external}.

The red and green coding of the crosswalk was also found to influence safety perception. Qualitative feedback suggested that the red crosswalk of interconnected eHMIs promoted cautious crossing behaviours among both groups of participants. Such induced feelings of stoppage and hesitation could reduce the likelihood of hasty decisions, affording both the pedestrian and the AV more time to react appropriately. However, continual discomfort may not be the optimal emotional state for urban environments. Echoing this concern, \citet{wang2022shared} emphasised the need for less obtrusive mechanisms that maintain pedestrian awareness without compromising their comfort and urban ambience.

\subsubsection{Workload} Quantitative results indicate that interconnected eHMIs did not lead to a reduction in cognitive load compared to unconnected eHMIs or the baseline condition. Intriguingly, neither type of eHMI showed a significant decrease in cognitive load compared to the absence of an eHMI. Although this observation might be attributed to pedestrians having to interpret additional visual signals, it contrasts with existing eHMI literature that often reports improved performance when cognitive load is a measured outcome, e.g., the blink LED light band~\cite{colley2022effects} or the smiley interface~\cite{prattico2021comparing}. A potential explanation for the differences observed in our results could be the traffic scenarios used in our study, especially the presence of traffic on a two-way street. In a similar study setting with a comparable projection-based concept, \citet{hollander2022take} also found that the workloads elicited in both the baseline and eHMI concept were closely matched. Furthermore, qualitative data indicates that both interconnected and unconnected eHMIs introduced distinct challenges, which may have contributed to consistently higher workloads.


The Unconnected condition had the lowest workload variance (92.43), suggesting a more homogeneous perception. This design offers a straightforward, non-dynamic response, which might be easier for a broader range of participants to comprehend and interpret. In contrast, the Interconnected condition had the highest variance (279.05), indicating that participants interpreted or experienced the concept in diverse ways. Qualitative data indeed showed that some participants might have found the dynamic changes and additional Wi-Fi symbols intuitive and helpful, while others might have viewed them as confusing or distracting. This variability in perceptions could arise from the between-subject factor, i.e., having knowledge of interconnected eHMI. However, given that the interaction between Group and Condition was not statistically significant, other factors like participants' tech familiarity or participants' capacity to process dynamic information quickly might be influential.

\subsubsection{Trust} 

According to the model of trust in automation~\cite{korber2019theoretical}, the perception of trustworthiness in automation hinges significantly on two factors: \textit{Reliability/Competence} and \textit{Understandability/Predictability}~\cite{korber2019theoretical}. Although another factor, \textit{The Intention of Developers} (i.e., car manufacturers), exists, we deemed it outside the scope of our research and therefore did not measure it. Our quantitative analysis has revealed that the interconnected eHMI scored marginally higher in these metrics compared to the Baseline and Unconnected alternatives. However, the difference did not reach statistical significance. A notable observation from qualitative analysis was the influence of the non-yielding scenario: 

\begin{itemize}
    \item Within the \textit{no knowledge} group, this scenario caused a discrepancy between what participants expected (or assumed) and what they actually observed in the scenario. Some participants correctly associated the Wi-Fi symbol with inter-vehicle communication. However, when they saw a vehicle continue without yielding despite the presence of the symbol, they abandoned this interpretation.
    \item For the \textit{knowledge} group, there is a prevailing expectation that if one AV stops for a pedestrian, subsequent AVs should follow suit. However, this cascade effect was observed inconsistently, leading to trust issues. This observation imply that a networked understanding and action among AVs is as important as accurately informing about traffic conditions.
\end{itemize}

Another observation that led to trust concerns was when all communication appeared to originate from a single AV, highlighted by just one participant. In their investigation of wearable AR as a solution for pedestrian interaction with AVs, \citet{tran2022designing} found that pedestrians preferred to have both individual AV responses and a clear signal to cross. In our study, using the interconnected eHMI, even though only a single vehicle provided information through the red/green coded crosswalk, other AVs displayed the abstract light signal. This might explain why only one participant raised this concern.

\subsection{Effect of Knowledge (RQ2) } 

Given the prior information on interconnected eHMIs, we would expect the \textit{knowledge} group to have a better grasp of this design concept. Surprisingly, regardless of the conditions, those in the \textit{knowledge} group perceived a significantly lower level of Understandability/Predictability compared to the \textit{no knowledge} group. The \textit{no knowledge} group might be making assumptions or relying on their intuition to gauge Understandability/Predictability, leading to an inflated sense of confidence. On the other hand, the \textit{knowledge} group, having received explicit information, might become more aware of the nuances and complexities, leading to a reduced Understandability/Predictability ratings.

Quantitative analysis suggests both groups had a similar responses when exposed to the interface conditions. However, the \textit{knowledge} group had a more favourable view of the Interconnected eHMI across all measures. Even though the interaction effects were not statistically significant, these trends could offer valuable insights for eHMI education (e.g. via public campaigns). In particular, the absence of an interaction effect suggests that the information alone does not uniquely alter the way users respond to different eHMI designs. This observation might be explained by the theory-practice gap. While reading materials provide a theoretical perspective, experiencing the concept in a scenario emphasises practice. For complex design concepts, education should ideally balance both to ensure a holistic understanding and effective user interaction with the system. 

This balance is evident in a study by \citet{faas2020longitudinal} where participants were introduced to eHMI concepts and directly experienced them during the first session. In the following sessions (7-9 days later), participants' recall of these concepts was tested, and if necessary, corrected before they experienced the concepts again. The authors discovered that the positive impacts of eHMIs not only persist over time in areas such as acceptance and user experience but may also intensify in others like trust and reliance. This underscores the importance of firsthand interaction and longitudinal assessment in understanding the true impact of eHMIs on users. 

\subsection{Practical Implications and Future Directions on eHMI Design} 

\textit{Balancing Trust and Overtrust:} Trust in automation is complex. While users need to trust the technologies they use, overtrust can lead to complacency and risky behaviours. Striking a delicate balance between clear guidance and user caution is crucial. For instance, \citet{faas2021calibrating} suggested teaching pedestrians to detect AV malfunctions via eHMI. Our red crosswalk solution aimed to instill vigilance, as seen in the study's observed crossing behaviour. However, feedback was mixed: half perceived this combination as a reinforced cautionary signal (interpreting red as \textit{caution}), while the other half found it conflicting (interpreting red as \textit{danger/stop}). In light of these divergent perceptions, future research might explore more nuanced design strategies. For instance, a less alarming colour, such as a darker shade of red typically seen in red coloured asphalt for road markings~\cite{mainroadsWA2016}, orange or yellow, could be considered to communicate warnings more subtly~\cite{wogalter2015color}. Enlarging the eHMI's crosswalk projection to span both lanes may clarify the extent of the covered space. Additionally, integrating symbols or text-based cues like \textit{`Cross with caution'} could further enhance clarity and accessibility, echoing the suggestions of some study participants. These suggestions align with the existing practice of using highly visible pavement markings to caution pedestrians\footnote{\url{https://www.federationcouncil.nsw.gov.au/Living-Here/Transport-Road-Safety/Look-Out-Before-You-Step-Out}, last accessed February 2024.}.

\textit{Communicating Traffic Intentions:} Traditional eHMIs primarily convey the intentions of individual vehicles, akin to a pedestrian seeking cues from a single driver. Interconnected eHMIs, however, offer insights into multiple vehicles. Our qualitative findings on collisions suggest that interconnected eHMIs can reduce signal misinterpretations, leading to safer interactions in traffic. This is particularly beneficial in complex traffic scenarios like multi-lane crossings or roundabouts, where the intentions of several vehicles must be interpreted cohesively.
The term interconnected eHMIs serves as an overarching concept encompassing various representations. While our study explored a centralised communication approach, future research might explore alternative representations. For example, while each AVs activates its eHMI individually, their yielding behaviour, as well as the timing and visual aspects of the communication signals (e.g., light band and projection), could be synchronised. This coordination would ensure temporal alignment and visual consistency in the communication across all AVs, enhancing safety and clarity in complex traffic situations.

\textit{Testing eHMIs in Mixed Traffic Environments:} In settings where vehicles, with and without eHMIs, coexist, eHMI designs should not complicate the pedestrian's decision-making process~\cite{mahadevan2019av}. Instead, they should complement the kinematic vehicle cues pedestrians have historically depended on. Given the potential challenges pedestrians face in mixed traffic situations, there is a pressing need for eHMI designers and researchers to undertake rigorous testing within these contexts, ensuring designs effectively enhance, rather than impede, pedestrian safety.

\subsection{Limitations}

Our design choices, such as the implementation of the red crosswalk and the Wi-Fi visual, while reasonable and informed, proved insufficient in facilitating a clear understanding of the interconnected eHMI operations among participants. As a result, the study findings pertain to this specific representation of interconnected eHMIs rather than the broader concept of interconnected eHMIs. This issue of generalisability is indeed a common challenge in eHMI design studies, where findings are often specific to the tested design. Despite these limitations, our study provides valuable benchmarks and insights for the future development of more effective eHMI systems, particularly in understanding user needs and preferences.

Our study assumed that participants' behaviour would be similar across designed scenarios. However, individual variations, influenced by past experiences, cognitive abilities, and risk tolerance, were significant. This variability was evident when some participants quickly crossed the street before the red crosswalk changed from red to green, potentially leading to an incomplete understanding of the interconnected eHMIs and their influence on measured effectiveness. Moreover, to simplify the study design and focus solely on interactions between AVs and pedestrians, all scenarios involved an AV stopping in the first lane. This decision resulted in a missed opportunity to compare pedestrian behaviour towards AVs and manually-driven vehicles in mixed traffic. Future work should aim to evaluate the interconnected eHMIs holistically under more varied situations.

Our reliance on a VR simulation introduced inherent limitations in terms of transferability to real-world scenarios, especially regarding field of view and the accuracy of estimating distance and speed. Additionally, the repetitive nature of having participants cross the street multiple times might not accurately reflect real-world pedestrian spontaneity. 

\section{Conclusion}

Our study underscores the potential benefits and challenges of integrating interconnected eHMIs into the urban landscape. Interconnected eHMIs significantly enhanced feelings of safety compared to the baseline, and they seemed to influence pedestrians to exercise more caution during crossings. The system also reduced the likelihood of signal misinterpretation when compared to unconnected eHMI system. However, scenarios without any eHMI reported the lowest number of collisions, suggesting a nuanced relationship between eHMIs and pedestrian safety, potentially influenced by the cognitive demands they introduce. The interpretation of signals from interconnected eHMIs emerges as a topic of interest from our findings, indicating the importance of clarity and consistency in design. Our data revealing that prior knowledge had no significant effect and, interestingly, appeared to diminish trust measures. As the landscape of urban mobility continues to evolve, the findings from this study contribute to the ongoing discourse on the integration of AVs and eHMIs, offering insights into the complexities inherent to this technology.

\begin{acks}
This research is supported by an Australian Government Research Training Program (RTP) Scholarship and through the ARC Discovery Project DP200102604, Trust and Safety in Autonomous Mobility Systems: A Human-centred Approach. The authors acknowledge the statistical assistance of Kathrin Schemann of the Sydney Informatics Hub, a Core Research Facility of the University of Sydney. We thank all the participants for taking part in this research. We extend our gratitude to the anonymous reviewers for their insightful comments and suggestions.
\end{acks}

\bibliographystyle{ACM-Reference-Format}
\bibliography{ACM/references}

\end{document}